\documentclass[letterpaper, 10 pt, conference]{ieeeconf}

\IEEEoverridecommandlockouts  

\usepackage[hidelinks]{hyperref}

\usepackage{fix-cm}
\usepackage{etex}

\usepackage{dblfloatfix}

\usepackage{nag}

\makeatletter
\@ifpackageloaded{xcolor}{}{%
\usepackage[table,x11names,dvipsnames,svgnames]{xcolor}%
}
\makeatother

\usepackage{colortbl}

\usepackage{graphicx}
\usepackage{wrapfig}

\definecolorset{RGB}{lyft}{}{Red,194,39,36;Sunset,202,53,33;Orange,205,68,20;Amber,200,117,42;Yellow,242,169,52;Citron,186,188,44;Lime,112,159,33;Green,56,139,31;Mint,45,118,56;Teal,52,133,135;Cyan,60,132,202;Blue,55,94,248;Indigo,64,13,247;Purple,115,42,248;Pink,176,25,145;Rose,176,32,75}

\usepackage{cite}

\usepackage{microtype}

\usepackage[american]{babel}

\usepackage{array}
\usepackage{multirow}
\usepackage{booktabs}
\usepackage{makecell} 

\ifcsname labelindent\endcsname
\let\labelindent\relax
\fi
\usepackage[inline]{enumitem}

\usepackage{subfig}

\newenvironment{lenumerate}[2][]
{\begin{enumerate}[label=(#2\arabic*),leftmargin=0.2in,itemindent=0.15in,#1]}
{\end{enumerate}}

\setlist*[enumerate,1]{label={\itshape\arabic*)}}

\makeatletter
\newcommand{\paragraphswithstop}{%
\let\copyparagraph\paragraph%
\renewcommand\paragraph[1]{\copyparagraph{##1.}}%
}
\makeatother

\usepackage[framemethod=tikz]{mdframed}

\makeatletter
\def\namedlabel#1#2{\begingroup
  #2%
  \def\@currentlabel{#2}%
  \phantomsection\label{#1}\endgroup
}
\makeatother

\makeatletter
\def\namedlabelphantom#1#2{\begingroup
  \def\@currentlabel{#2}%
  \phantomsection\label{#1}\endgroup
}
\makeatother

\newcommand{\parunskip}{\bgroup\unskip\parfillskip=0pt \par\egroup}

\input{preamble/math}
\newcommand{\real}[1]{\mathbb{R}^{#1}{}}

\newcommand{\bmat}[1]{\begin{bmatrix}#1\end{bmatrix}}

\newcommand{\transpose}{^\mathrm{T}}

\DeclarePairedDelimiter{\abs}{\lvert}{\rvert}

\newcommand{\vct}[1]{\mathbf{#1}}

\DeclareMathOperator*{\argmin}{\arg\!\min}

\DeclareMathOperator{\stack}{stack}

\newcommand{\union}{\cup}

\newcommand{\subjectto}{\textrm{subject to}\;}

\providecommand{\va}{\vct{a}}

\providecommand{\vb}{\vct{b}}

\providecommand{\vx}{\vct{x}}

\providecommand{\vy}{\vct{y}}

\providecommand{\mA}{\vct{A}}

\providecommand{\mK}{\vct{K}}

\providecommand{\mS}{\vct{S}}

\providecommand{\cC}{\mathcal{C}}

\providecommand{\cF}{\mathcal{F}}

\providecommand{\cK}{\mathcal{K}}
\providecommand{\cL}{\mathcal{L}}

\providecommand{\cX}{\mathcal{X}}

\usepackage{units}

  \newcommand{\newcolorlabel}[2]{%
  \expandafter\newcommand\csname #1\endcsname[1]{%
    \tikz[baseline]{\node[text=white,fill=#2,anchor=base,text height=1.3ex,text depth=0.1ex,font=\sffamily\bfseries]{##1}}}%
}

\newcommand{\newcommenter}[2]{%
  \expandafter\newcommand\csname #1\endcsname[1]{%
    \fcolorbox{#2}{#2}{\color{white}\textsf{\textbf{#1}}}
    {\color{#2}##1}}%
  \expandafter\newcommand\csname at#1\endcsname{%
    \fcolorbox{#2}{#2}{\color{white}\textsf{\textbf{@#1}}}
    {\color{#2}}}%
  \expandafter\newcommand\csname #1cite\endcsname[1]{%
    \csname #1\endcsname {[##1]}
  }%
  \expandafter\newcommand\csname #1ref\endcsname[1]{%
    \csname #1\endcsname {$\blacktriangleright$##1}
  }%
  \expandafter\newcommand\csname #1hl\endcsname[2]{%
    \colorbox{#2}{\color{white}\textsf{\textbf{#1}}}\sethlcolor{Azure2}\hl{##2}~%
    \expandafter\ifx\csname commentarrow\endcsname\relax$\leftarrow$\else \commentarrow[#2]\fi~%
    {\color{#2}##1}}%
  \expandafter\newcommand\csname #1st\endcsname[2]{%
    \colorbox{#2}{\color{white}\textsf{\textbf{#1}}}\sout{##2}~%
    \expandafter\ifx\csname commentarrow\endcsname\relax$\leftarrow$\else \commentarrow[#2]\fi~%
    {\color{#2}##1}}%
}
\newcommenter{TODO}{DodgerBlue1}
\newcommenter{rtron}{Green3}

\usepackage{comment}

\usepackage{pdfcomment}

\usepackage{soul}

\usepackage[normalem]{ulem}

\usepackage{csquotes}

\usepackage{suffix}

\usepackage{environ}

\makeatletter
\newsavebox{\boxifnotempty}
\newcommand{\displayifnotempty}[3]{\sbox\boxifnotempty{#2}\setbox0=\hbox{\usebox{\boxifnotempty}\unskip}%
  \ifdim\wd0=0pt
  \else
  #1\usebox{\boxifnotempty}#3%
  \fi%
}

\newcommand{\ifempty}[2]{\setbox0=\hbox{#1\unskip}%
  \ifdim\wd0=0pt%
  #2%
  \fi%
}

\newcommand{\ifnotempty}[2]{\setbox0=\hbox{#1\unskip}%
  \ifdim\wd0>0pt%
  #2%
  \fi%
}
\makeatother

\newcommand{\switchifempty}[3]{\sbox\boxifnotempty{#1}\setbox0=\hbox{\usebox{\boxifnotempty}\unskip}%
  \ifdim\wd0=0pt{}%
  #2%
  \else{}%
  #3%
  \usebox{\boxifnotempty}%
  \fi%
}

\makeatletter
\@ifundefined{chapter}{\usepackage{algorithm}}{\usepackage[chapter]{algorithm}}
\makeatother
\usepackage{algorithmicx}
\usepackage{algpseudocode}
\makeatother%

\usepackage{scrlfile}

\makeatletter
\newcommand*\newstoreddef[1]{
  \BeforeClosingMainAux{%
    \immediate\write\@auxout{%
      \string\restoredef{#1}{\csname #1\endcsname}%
    }%
  }%
}
\newcommand*{\restoredef}[2]{
  \expandafter\gdef\csname stored@#1\endcsname{#2}%
}
\newcommand*{\storeddef}[1]{
  \@ifundefined{stored@#1}{0}{\csname stored@#1\endcsname}%
}
\makeatother

\usepackage{pageslts}
\NewEnviron{tee}{\BODY\typeout{Marker Tee [start] ^^J \BODY ^^JMaker Tee [end]}}

\usepackage{cleveref}

\usepackage{tikz}
\usetikzlibrary{calc}
\usetikzlibrary{matrix}
\usetikzlibrary{chains,scopes}
\usetikzlibrary{shapes.geometric}
\usetikzlibrary{arrows.meta}
\usetikzlibrary{decorations.markings}
\usetikzlibrary{decorations.pathreplacing}
\usetikzlibrary{backgrounds}

\tikzset{
  dim above/.style={to path={\pgfextra{
        \pgfinterruptpath
        \draw[>=latex,|->|] let
        \p1=($(\tikztostart)!1.5em!90:(\tikztotarget)$),
        \p2=($(\tikztotarget)!1.5em!-90:(\tikztostart)$)
        in(\p1) -- (\p2) node[pos=.5,sloped,above]{#1};
        \endpgfinterruptpath
      }
    }
  },
  dim double above/.style={to path={\pgfextra{
        \pgfinterruptpath
        \draw[>=latex,|->|] let
        \p1=($(\tikztostart)!3em!90:(\tikztotarget)$),
        \p2=($(\tikztotarget)!3em!-90:(\tikztostart)$)
        in(\p1) -- (\p2) node[pos=.5,sloped,above]{#1};
        \endpgfinterruptpath
      }
    }
  },
  dim below/.style={to path={\pgfextra{
        \pgfinterruptpath
        \draw[>=latex,|->|] let 
        \p1=($(\tikztostart)!-1em!-90:(\tikztotarget)$),
        \p2=($(\tikztotarget)!-1em!90:(\tikztostart)$)
        in (\p1) -- (\p2) node[pos=.5,sloped,below]{#1};
        \endpgfinterruptpath
      }
    }
  },
}

\tikzset{
    right angle quadrant/.code={
        \pgfmathsetmacro\quadranta{{1,1,-1,-1}[#1-1]}     
        \pgfmathsetmacro\quadrantb{{1,-1,-1,1}[#1-1]}},
    right angle quadrant=1, 
    right angle length/.code={\def\rightanglelength{#1}},   
    right angle length=2ex, 
    right angle symbol/.style n args={3}{
        insert path={
            let \p0 = ($(#1)!(#3)!(#2)$) in     
                let \p1 = ($(\p0)!\quadranta*\rightanglelength!(#3)$), 
                \p2 = ($(\p0)!\quadrantb*\rightanglelength!(#2)$) in 
                let \p3 = ($(\p1)+(\p2)-(\p0)$) in  
            (\p1) -- (\p3) -- (\p2)
        }
    }
}

\newcommand{\pgfextractangle}[3]{%
    \pgfmathanglebetweenpoints{\pgfpointanchor{#2}{center}}
                              {\pgfpointanchor{#3}{center}}
    \global\let#1\pgfmathresult  
}

\usetikzlibrary{shapes.arrows}
\newcommand{\commentarrow}[1][Azure4]{\tikz[baseline=-3pt]{\node[shape border uses incircle, fill=#1,rotate=180,single arrow, inner sep=1pt, minimum size=6pt, single arrow head extend=2pt]{};}}

\tikzset{ax/.style={-latex,line width=2pt}}

\tikzset{camera/.style={fill=Sienna1,fill opacity=0.5},%
image plane/.style={draw=RoyalBlue3,line width=2pt}}

\makeatletter
\newcommand{\@LN@col}[1]{}
\newcommand{\@LN}[2]{}
\makeatother
\newcommenter{todo}{Firebrick1}
\newcommenter{mmitjans}{Magenta}
\newcommenter{tomwu}{Blue}

\usepackage[font=small,labelfont=bf]{caption}

\renewcommand{\thetable}{\arabic{table}} 

\title{\LARGE \bf
Learning deep Koopman operators with convex stability constraints
}

\author{Marc Mitjans$^{1}$, Liangting Wu$^{1}$ and Roberto Tron$^{1,2}$
\thanks{This work was not supported by any organization.}
\thanks{$^{1}$Marc Mitjans, Liangting Wu and Roberto Tron are with the Department of Mechanical Engineering,
        Boston University, 110 Cummington Mall, MA 02215, United States
        {\tt\small mmitjans@bu.edu, tomwu@bu.edu, tron@bu.edu}.}%
\thanks{$^{2}$Roberto Tron is also with the Division of Systems Engineering, Boston University, 44 Cummington Mall, MA 02215, United States.
}%
}

\begin{document}

\maketitle
\thispagestyle{empty}
\pagestyle{empty}

\begin{abstract}
In this paper, we present a novel sufficient condition for the stability of discrete-time linear systems that can be represented as a set of piecewise linear constraints, which make them suitable for quadratic programming optimization problems. More specifically, we tackle the problem of imposing asymptotic stability to a Koopman matrix learned from data during iterative gradient descent optimization processes. We show that this sufficient condition can be decoupled by rows of the system matrix, and propose a control barrier function-based projected gradient descent to enforce gradual evolution towards the stability set by running an optimization-in-the-loop during the iterative learning process. We compare the performance of our algorithm with other two recent approaches in the literature, and show that we get close to state-of-the-art performance while providing the added flexibility of allowing the optimization problem to be further customized for specific applications.

\end{abstract}


\section{INTRODUCTION}
Koopman operator is a powerful tool for investigating nonlinear dynamical systems and has gained increasing interests over the last two decades in areas such as system identification \cite{mauroy2016linear,bruder2019nonlinear} and nonlinear control \cite{korda2018linear}. It is an infinite dimensional linear operator that acts on a set of observable functions of the original system states \cite{mauroy2020koopman}. 
To circumvent the infinite dimension property and assessing the operator in a practical way, started from \cite{schmid2010dynamic, rowley2009spectral}, numerous data driven methods are developed. These methods find the so-called Koopman modes from data and computes a finite-dimensional approximation of the Koopman operator. 

One well-known method that finds Koopman operators from data is the Extended Dynamic Mode Decomposition (EDMD) algorithm \cite{williams2015data}, which is based on the DMD algorithm, a powerful technique for analyzing observative data of high-dimensional complex systems \cite{proctor2016dynamic}. DMD efficiently finds the dynamics of the system states by computing the eigenmodes and eigenvalues of the system. Extended Dynamic Mode Decomposition (EDMD) improves DMD by constructing a dictionary of observable functions of the states before passing it to the DMD algorithm. In \cite{korda2018convergence}, the convergence behavior of EDMD as the number of samples goes to infinity is investigated. 

The performance of EDMD highly depends on the selection of the observables, thus requires good measurements of the system. To overcome this issue, deep learning is used to learn the observables and the Koopman operator simultaneously \cite{li2017extended, lusch2018deep, azencot2020forecasting, yeung2019learning}. The Deep-DMD method is flexible to different systems without the need of building dictionaries in advance, outperforming the EDMD methods with fewer observables yet higher accuracy \cite{yeung2019learning }. 

The application of the Koopman operator in control has been extensively studied in recent years, e.g., \cite{ williams2016extending, proctor2016dynamic, proctor2018generalizing, li2019learning, abraham2019active}. As a linear embedding of nonlinear systems, the Koopman operator enables the use of linear control tools and can sometimes exhibit better performance than some nonlinear control policies \cite{kaiser2021data}. Efforts have been made to apply the Koopman operator in the design of controllers such as Model Predictive Control \cite{korda2018linear, kaiser2018sparse}. 

For the purpose of improving long-term prediction performance, Koopman operators with stability guarantees is an active field of research. The work in \cite{ mamakoukas2023learning} shows that stability constraints can improve the overall predictive accuracy and provides conditions for choosing observables that lead to system stability. It is shown in \cite{ yi2023equivalence} that the stability in the Koopman framework implies the existence of a contraction metric of the nonlinear system. In \cite{fan2022learning}, a new Koopman learning framework is built and,  by parameterization, the Koopman matrix is Schur stable. Similar to \cite{fan2022learning}, another Koopman learning framework that ensures stability is developed in \cite{ bevanda2022diffeomorphically}, but for continuous time systems.

In this paper, we further delve into the research on imposing stability to the Koopman operator, providing the following contributions:
\begin{itemize}
    \item We introduce a sufficient stability condition for discrete-time linear dynamical systems that is piecewise linear with respect to the system matrix parameters, in contrast to general algebraic conditions (such as the matrix being Schur-stable).
    \item The sufficient stability condition can be decoupled by the rows of the system matrix. Together with their piecewise linearity, they become suitable as constraints for quadratic programming (QP) optimization problems.
    \item We implement this condition by proposing a control barrier function (CBF)-based projected gradient descent (PGD) to enforce eventually meeting the stability condition during the training of a Koopman system while avoiding abrupt undesired updates of the parameter.
\end{itemize}

\section{PROBLEM OVERVIEW AND PRELIMINARIES}

In this section, we present the mathematical background and foundational concepts that lead to our method for stability guarantees defined in \cref{sec:cbf-based-pdg} of this paper. 

\subsection{The Koopman operator}

Consider the following discrete-time dynamical system:
\begin{equation}
    \vx^+ = f(\vx),
    \label{eq:general-discrete-system}
\end{equation}
where $\vx \subset \cX \in \real{n}$. Koopman theory states that there is a higher dimensional vector-space $\cF$ in which the same dynamics evolve in a linear manner:
\begin{equation}
    (\cK\psi)(\vx) = (\psi \circ f)(\vx) = \psi(\mathbf{x}^+),
    \label{eq:koopman-eq}
\end{equation}
where $\psi(\vx) \subset \cF \in \real{d}$, for $d>n$, are called \emph{observable functions} of the system and describe a linear evolution of the system in the so-called \emph{lifted} space $\cF$ \cite{mauroy2020koopman}. 
However, since this linear operator acts on continuous functions of the state $\vx$, it is, in general, infinite dimensional. In practice, data-driven methods can be used to approximate the infinite-dimensional operator $K$ by a finite dimensional matrix $\mK$.




\subsection{Data-driven approximations of the Koopman operator}

There are multiple state-of-the-art techniques in the literature that approximate the Koopman operator with data, the most well-known being the Extended Dynamic Mode Decomposition (EDMD). EDMD provides a closed-loop solution to the Koopman operator by solving an unconstrained least-squares optimization problem. Consider pairs of consecutive system states $\left\{\mathbf{x}_k,\mathbf{y}_k\right\}$, where $\vy_k = f(\vx_k)$, and let $\psi$ be a vector of $d$ observable functions $\psi=\stack(\psi_1,\dots,\psi_d)$. 
Let us define the matrices
\begin{equation}
    \begin{aligned}
        \Psi^+=\bmat{\ldots & \psi(\vy_k) & \ldots}, && \Psi=\bmat{\ldots & \psi(\vx_k) & \ldots}
    \end{aligned}
\end{equation}
by concatenating alongside the column dimension all the column-vector observables $\psi_k$.
Then, the approximate Koopman system can be estimated from data by solving following least-squares problem:
\begin{equation}
    \mK^* = \argmin_{\mathbf{K}} \| \Psi^+ - \mathbf{K} \Psi  \|_2,
    \label{eq:edmd}
\end{equation}
which has the closed-form solution
\begin{equation}
    \mathbf{K}^* =  \Psi^+ \Psi^\dagger,
\end{equation}
where $\Psi^\dagger$ corresponds to the Moore-Penrose pseudoinverse of $\Psi$.
Note, however, that the above method has some evident limitations:
\begin{enumerate}
    \item This method assumes the dictionary of observables is handpicked beforehand, rather than learning it simultaneously with the Koopman matrix $\mathbf{K}$.
    \label{limitation-one}
    \item The closed-form solution of this method doesn't allow for additional objective functions (such as multi-step prediction objectives). 
    \label{limitation-two}
    \item This method does not consider the stability of the approximate Koopman matrix $\mK$.
    \label{limitation-three}

    
\end{enumerate}

Recent studies \cite{yeung2019learning,li2022koopman} have demonstrated that neural networks can enhance the efficiency of the dictionary of observables by jointly learning both the lifting functions and the Koopman matrix from data, thus directly addressing both limitations \ref{limitation-one} and \ref{limitation-two}. We tackle limitation \ref{limitation-three} by applying the control barrier function technique to a new sufficient condition for stability presented in \cref{sec:stability-of-inner-point} and implemented
in a gradient descent optimization problem in sections \ref{sec:cbf-based-pdg}--\ref{sec:numerical-analysis}.

\subsection{Control barrier functions}
\label{sec:cbf-formulation}

Consider the dynamical system \eqref{eq:general-discrete-system}, a feasible set $\cC\in\cX$ where $\vx$ is allowed to evolve, and its complementary infeasible set $\bar{\cC}$.
A control barrier function (CBF) is defined in the context of dynamical systems as a function $h(\vx)$ such that $h>0$ inside $\cC$, $h<0$ inside $\bar{\cC}$, and $h(\vx) = 0$ at the boundary. 
\begin{lemma}[based on \cite{agrawal2017discrete}]
If the constraint
\begin{equation}
    h(\vx_{k+1}) \geq \alpha\bigl(h(\vx_k)\bigr)
\end{equation}
is satisfied $\forall k$ with  $\alpha$ being a class-$K$ function, 
the system $f(x)$ is guaranteed to never enter the infeasible set $\bar{\cC}$ if $h(\vx_0) \geq 0$, and to asymptotically approach the feasible set if $h(\vx_0) < 0$.
\end{lemma}
\subsection{A novel set-based stability condition for discrete-time linear systems}
\label{sec:stability-of-inner-point}


In this section, we present a new sufficient stability condition for linear discrete-time systems that can be expressed using linear constraints (as opposed to, for instance, SDP constraints \cite{boyd1994linear}).
Let $\cC:=\{\vx|\mA_{c}\vx\leq \vb_c\}$ be a polyhedral set that contains the origin, i.e., such that $0\in\cC$. Note that, by substituting the origin in the definition of $\cC$, we have 
\begin{equation}\label{eq:b positive}
    \vb_c\geq 0.
\end{equation}
Given a non-negative number $s\geq 0$, we define the \emph{scaled set} $s\cC$ as $s\cC=\{s\vx | \vx\in \cC\}$. 
\begin{lemma}
The scaled set $s\cC$, $s\geq 0$ can be equivalently defined as
\begin{equation}
    s\cC=\{\vx'|\mA_c\vx'\leq s\vb_c\}
\end{equation}
\end{lemma}
\begin{proof}
The claim follows by using the change of variable $\vx'=s\vx$ and the definition of $\cC$.
\end{proof}
Scaled sets containing the origin have the following \emph{nesting property}:
\begin{corollary}\label{corrollary:nesting}
    We have that $s'\cC\subseteq s\cC$ if and only if $s'\leq s$.
\end{corollary}
\begin{proof}
If $\vx\in s'\cC$, then $\mA_c\vx\leq s'\vb_c\leq s\vb_c$ (since $\vb_c$ is positive by \eqref{eq:b positive}), hence $\vx\in s\cC$. The other direction of the claim can be proved by picking an arbitrary $\vx\neq 0$ and following a similar argument.
\end{proof}
For the set $\cC$ and a discrete time system of the form
\begin{equation}\label{eq:discrete system}
    \vx^+=\mA\vx,
\end{equation}
we proceed by making the following definitions.
\begin{definition}
A field $\mA\vx$ is said to be \emph{inward-pointing} with respect to $\cC$ if $\mA\vx_0\in\cC$ for all $\vx_0\in\partial\cC$.
\end{definition}
\begin{corollary}\label{def:definition_1}
Let $\va_{cj}\transpose$, $b_{cj}$ denote the $j$-th row of $\mA_c$ and $\vb_c$, respectively. 
A vector field is inward-pointing with respect to $\cC$ if and only if 
\begin{equation}
\va_{cj}\transpose\mA \vx_0 \leq b_{cj}\label{eq:pointwise def inward pointing}
\end{equation}
for all $j$ corresponding to constraints that are active at $\vx_0$, i.e., for all $j$ that satisfy 
\begin{equation}
    \va_{cj}\transpose \vx_0=b_{cj}.
\end{equation}
\end{corollary}
\begin{proof}
This equivalent definition follows from the fact that $\vx_0\in\cC$ iff one or more constraints are active.
\end{proof}

We now combine the notion of inward-pointing with the nesting property.
\begin{lemma}\label{lemma:inward scaled}
Assume the set $\cC$ contains the origin, i.e., $0\in\cC$. A field $\mA\vx$ is inward-pointing with respect to $\cC$ if and only if it is inward-pointing for any $s\cC$, $s>0$.
\end{lemma}
\begin{proof}
A point $\vx$ is on the boundary of $\cC$ if and only if the corresponding point $x'=s\vx$ will be on the boundary of $s\cC$, i.e.,
\begin{equation}\label{eq:boudary scaling}
 \vx\in\partial \cC \iff \vx'\in\partial s\cC
\end{equation}
(just multiply by $s$ both sides of $\mA_c\vx\leq \vb_c$ in the definition of $\cC$).
Let ${\vx'}^+=A\vx'$.
Multiplying both sides of \eqref{eq:pointwise def inward pointing} by $s$ we then have that \eqref{eq:boudary scaling} holds if and only if the following holds
\begin{equation}
\vx^+\in\cC\iff {\vx'}^+\in s\cC.
\end{equation}
The claim follows.
\end{proof}
This property of nested scaled sets with inward-pointing fields implies the following.
\begin{proposition}\label{proposition:forward invariance}
    Assume the dynamics \eqref{eq:discrete system} is inward-pointing with respect to $\cC$. Then $\cC$ is forward-invariant.
\end{proposition}
Intuitively, this proposition says that the linearity of the field allows us to extend the inward-pointing property to any scaled version of the set $\cC$ (for scales both smaller and larger than one).

\begin{proof}
To claim forward invariance, we need $\mA\vx\in\cC$ for all $\vx\in\cC$. If $\vx$ is on the boundary, we are done thanks to the definition of inward-pointing. If $\vx$ is in the interior of $\cC$, we proceed by contradiction. Assume that $\vx\in\cC$ but $\mA\vx\notin \cC$. Let $s$ be the minimum scale such that  containing $\vx\in s\cC$. We have $s<1$ (otherwise, by \Cref{corrollary:nesting}, we would not have that $\vx$ is in the interior of $\cC$). This would mean that $s\cC$ is not inward-pointing, leading to a contradiction with \Cref{lemma:inward scaled}. We therefore conclude that $x^+\in s\cC \subseteq \cC$, hence $\cC$ is forward invariant.
\end{proof}
\Cref{proposition:forward invariance} implies stability of the linear dynamics \eqref{eq:discrete system}.
\begin{theorem}
    If the dynamics \eqref{eq:discrete system} is inward-pointing with respect to any non-degenerate polyhedral set $\cC$, then it is stable with an equilibrium at $x_{eq}=0$.
    \label{theorem:inward-pointing}
\end{theorem}
\begin{proof}
The origin is always an equilibrium (since $\mA 0=0$). For any arbitrary point $\vx$ we can always find an $s>0$ such that $\vx\in s\cC$. Then, from \Cref{lemma:inward scaled} and \Cref{proposition:forward invariance} we have that the trajectory generated starting from $\vx$ will always stay in $s\cC$. Hence the system is stable.
\end{proof}

Theorem \ref{proposition:forward invariance} provides a visual representation of a sufficient condition for asymptotic stability.
In the next section, we start from this proposition to develop a set of piecewise linear constraints to impose stability onto a learnable Koopman matrix $\mK$.

\section{CBF-BASED PGD FOR STABILITY GUARANTEES}
\label{sec:cbf-based-pdg}

Given the sufficient condition for asymptotic stability suggested in \cref{theorem:inward-pointing}, we propose a constrained optimization formulation for projected gradient descent based on control barrier functions that imposes stability on a learned Koopman matrix $\mK\in \real{d\times d}$. The novelty of this approach is that the stability condition is piecewise linear and can decoupled into rows of $\mK$, thus reducing the number of variables of each optimization problem by a factor of $d$. Interestingly, this CBF formulation for iterative gradient descent
can also be generalized beyond the stability of a linear dynamical system, and applied to problems that may require other hard constraints on model parameters.

Without loss of generality, consider an equilibrium point  $\psi_{eq} = \mathbf{0}$ of a discrete-time Koopman system described as $\psi^+ = \mK\psi$ (for $\psi_{eq} \neq \mathbf{0}$, one can apply a simple change of coordinates to place the equilibrium point at $\mathbf{0}$). 
Moreover, consider a polyhedron $\cC$ defined as the axis-aligned hypercube $\cC=[-1,1]^d$.



The axis-alignment property of $\cC$ allows us to express the inward-pointing condition as follows.

\begin{lemma}
Let $\mK_{ij}$ be an element of the Koopman matrix $\mK$ corresponding to the $i$-th row and $j$-th column. Following \cref{theorem:inward-pointing}, the system $\psi^+ = \mK \psi$ is asymptotically stable if 
\begin{equation}
1\pm \mK_{ii}-\sum_{j\neq i} \abs{\mK_{ij}}\geq 0.
    \label{eq:lemma-unit-hypercube}
\end{equation}
\label{lemma:unit-hypercube}
\end{lemma}

\begin{proof}
Let $\cF_i$ (respectively, $\cF_{-i}$) be the face of $\cC$ where $\psi_i=1$ (respectively, $\psi_i=-1$). 
For our choice of $\cC$, we can check that $K\psi$ is inner pointing by projecting along each axis, obtaining  
\begin{align}
    -1\leq \sum_{i} K_{ij}\psi_j \leq 1 && \forall \psi \in \cF_i \union \cF_{-i}.
\end{align}

Using the facts that $\psi_i$ is fixed to $1$ and $-1$ on, respectively, $\cF_i$ and $\cF_{-i}$, $-1\leq\psi_j\leq1$ if $i\neq j$, and removing redundant inequalities, we have
\begin{equation}
\pm \mK_{ii} +\sum_j\min_{-1\leq \psi \leq 1} \mK_{ij} \geq -1.
\end{equation}
Substituting each minimization with the absolute value, the claim follows.
\end{proof}


Lemma \ref{lemma:unit-hypercube} provides the interesting sufficient condition for the asymptotic stability of $\mK$ that can be described as a series of convex and piecewise linear hard constraints decoupled by rows of $\mK$, rather than constraints affecting the entire matrix. Additionally, the convexity of these constraints make them suitable for quadratic programming (QP) optimization problems that seek to optimize over $\mK$, such as in constrained least-squares EDMD or in projection optimization problems.

\subsection{CBF-based projected gradient descent with stability constraints}
\label{subsec:cbf-based-pgd}

Consider the quadratic programming projection optimization problem defined as

\begin{problem}\label{problem:qp-projection-optimization}
\begin{equation}
\begin{aligned}
    &\min_{\mK} && \|\mK - \tilde{\mK} \|_2^2\\
    &\subjectto && \eqref{eq:lemma-unit-hypercube} \; \forall i,
\end{aligned}
\end{equation}
\end{problem}
\vspace{2mm}
where $\tilde{\mK}$ is the reference matrix. Then, the solution $\mK^*$ to problem \ref{problem:qp-projection-optimization} will guarantee stability. Such an optimization problem is usually solved within the broader scope of iterative gradient descent optimization problems, where certain model parameters need to satisfy certain hard constraints, and thus acts as the \emph{projected gradient descent} (PGD) step.


Consider the general optimizer-agnostic update rule of a Koopman matrix parameter $\mK$ applied at each learning step:
%
\begin{equation}
    \mK^+ = f(\mK,\gamma, \nabla_{\mK}\cL),
    \label{eq:k-update-dynamical-system}
\end{equation}
where $\gamma$ is the learning rate and $\nabla_{\mK}\cL$ is the gradient of the loss function $\cL$ to be optimized over with respect to $\mK$. This rule can be viewed as a discrete-time dynamical system, in which \cref{eq:lemma-unit-hypercube} acts as a control barrier function for each row $i$ of $\mK$:
\begin{equation}
    h_i(\mK) = 1 \pm \mK_{ii} - \sum_{\forall j \neq i} | \mK_{ij}| \geq 0.
\end{equation}

This allows us to re-write problem \ref{problem:qp-projection-optimization} as
\begin{problem}
\label{problem:qp-projection-optimization-rewritten}
\begin{equation}
\begin{aligned}
    &\min_{\mK^+} && \|\mK^+ - \tilde{\mK}^+ \|_2^2\\
    &\subjectto && h_i(\mK^+) \geq \min(0, \alpha h_i(\mK)) \; \forall i,
\end{aligned}
\end{equation}
\end{problem}
\vspace{2mm}
where $\mK$ is the pre-step update value of the parameter, and $\tilde{\mK}^+$ is the post-step update value that acts as a reference matrix for the PGD.
Problem \ref{problem:qp-projection-optimization-rewritten} adds the interesting property that, if $h_i(\mK)$ is not satisfied at the beginning of the training, the PGD step will not compute harsh updates for the parameter to guarantee feasibility, but instead will enforce that the parameter gradually evolves towards the feasible set in future updates. The $\min$ operator acts as a relaxation to the original CBF formulation: If the value before the step update already lies in the feasible set, it ensures that the CBF boundary still remains at $0$, preventing $h_i(\mK)$ from monotonically increasing.

Problem \ref{problem:qp-projection-optimization-rewritten} has a total of $d^2$ variables. However, considering that the squared L2-norm can be decoupled element-wise, and that
the stability condition established by \Cref{lemma:unit-hypercube} is applied to each row independently, the QP optimization problem can be segmented into $d$ optimizations of $d$ variables each. {This segmentation can be particularly beneficial when the dimension $d$ of the lifted space is substantially higher than that of the original space, as it simplifies a high-dimensional optimization into $d$ more manageable, lower-dimensional problems.}

Additionally, note that the $\pm \mK_{ii}$ term in the CBF is required  for those cases in which the data is sampled at very low rates and/or behaves erratically. In practice, however, it could be possible to drop the constraint containing the term $-\mK_{ii}$ if the data is observed to evolve smoothly.

\subsection{Expanding the set of learnable dynamics}

Since the sufficient condition presented by \cref{lemma:unit-hypercube} is dependent on the selected polyhedron, it becomes clear that this choice limits the set of dynamical systems that can fit the constraints in the PGD QP optimization.
To expand the set of learnable dynamics under these constraints, we introduce an additional learnable matrix $\mS$ that serves as a change of basis for $\mK$. Therefore, the Koopman dynamical system that will be learned is characterized by the following equation:
\begin{equation}
    \psi^+ = \mS^{-1} \mK \mS \psi.
\end{equation}

\begin{figure}[t]
    \centering
    \subfloat[Original system]{%
        \includegraphics[width=0.46\linewidth]{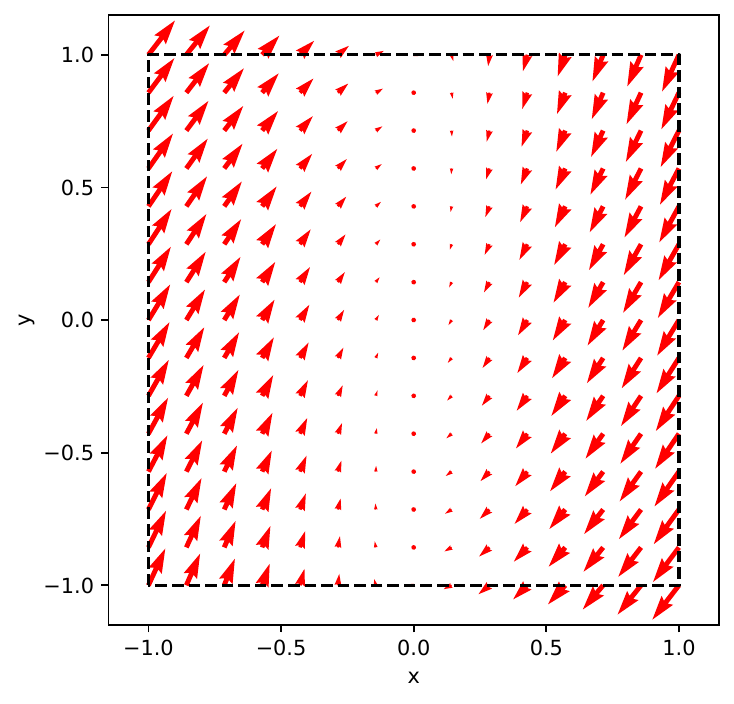}
        \label{fig:original-system}
    }
    \subfloat[Transformed system]{%
        \includegraphics[width=0.46\linewidth]{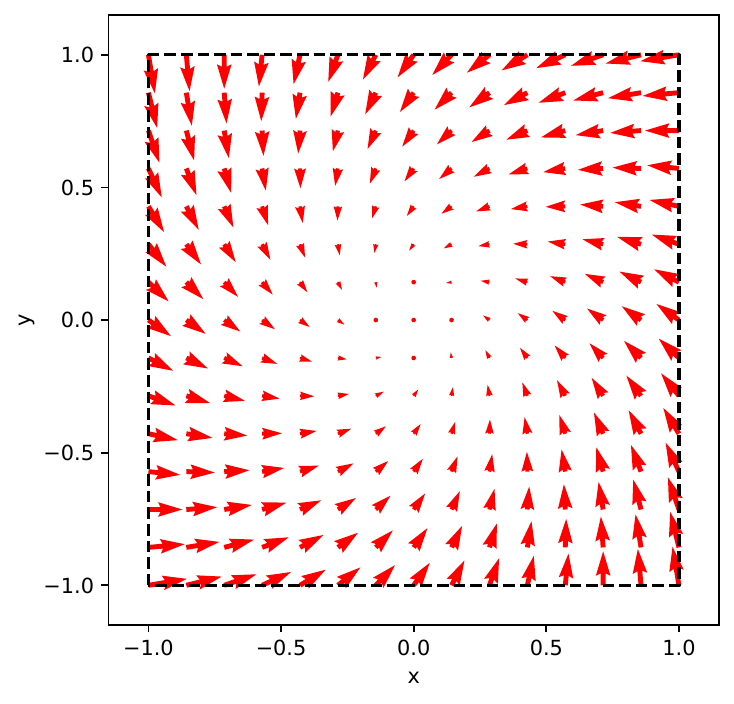}
        \label{fig:transformed-system}
    }
    \caption{Comparison of the vector fields of the same system in two different coordinate frames, where the bounding hypercube is depicted in dashed black lines. On the left, the original system clearly does not satisfy the constraints imposed by \cref{lemma:unit-hypercube}. On the right, after applying the right change of basis $\mS$, the same system is now compliant with the constraints.}
    \label{fig:vector-fields}
\end{figure}

\Cref{fig:vector-fields} shows how the change of basis matrix $\mS$ transforms an \emph{a priori} unlearnable dynamical system into one that does satisfy the constraints introduced by \Cref{lemma:unit-hypercube}.







\section{KOOPMAN LEARNING FRAMEWORK}
\label{sec:koopman-learning-framework}

We consider the problem of finding an approximated Koopman representation of some given nonlinear dynamics of the form of \cref{eq:general-discrete-system}, in which the observable functions $\psi(\vx)$ are parameterized with a neural network such that \mbox{$\psi(\vx) = \psi_{e}(\vx; \theta_{e})$}. We also define an additional neural network that acts as a decoder to recover the states in the original space given a Koopman observable: $\bar{\vx} = \psi_d(\psi(\vx); \theta_{d})$.

Then, the optimization problem of fitting a Koopman model to the data can be described as
\begin{problem}
\begin{equation}
    \mK^*, \theta_{e}^*, \theta_d^*, \mS^* = \argmin_{\mK, \theta_{e}, \theta_d, \mS} \cL,
\end{equation}
\end{problem}
\vspace{2mm}

where $\cL$ corresponds to the problem's loss function. We decompose $\cL$ into three distinct loss functions, each targeting a specific objective:
\begin{itemize}
    \item $\cL_{pred} = \sum_{k=1}^K \| \vx_k - \psi_d(\mK'^k \psi_e(\vx_0) \|_2$: This function targets minimizing the prediction error in the original state space along $K$ steps.
    \item $\cL_{lin} = \sum_{k=1}^K \| \psi_e(\vx_k) - \mK'^k \psi_e(\vx_0) \|_2$: This function minimizes the prediction error in the lifted Koopman space, enforcing the linearity condition.
    \item $\cL_{rec} = \sum_{k=1}^K \| \vx_k -\psi_d(\psi_e(\vx_k)) \|_2$: This function minimizes the reconstruction error of the original states by treating the composition of $\psi_e$ and $\psi_d$ analogously to an autoencoder,
\end{itemize}
where $\mK' = \mS^{-1} \mK \mS$.
Then, the overall loss function $\cL$ can be expressed as
\begin{equation}
    \cL = \lambda \cL_{pred} + \mu \cL_{lin} + \nu \cL_{rec}.
\end{equation}

$\cL$ can be minimized using an off-the-shelf unconstrained optimization toolbox (such as Pytorch \cite{paszke2017automatic}), which renders local optima for the system's parameters conditioned to the given set of hyperparameters and initial conditions. 
Then, the CBF-based PGD optimization defined in problem \ref{problem:qp-projection-optimization-rewritten} can be applied seamlessly after each update of $\mK$ for each row of the matrix to ensure that the updates lead the parameter towards the feasible set defined by \cref{lemma:unit-hypercube}, thus eventually reaching asymptotic stability.

\section{NUMERICAL PERFORMANCE EVALUATION}
\label{sec:numerical-analysis}

We validated our CBF-based projected gradient descent method to learn stable Koopman systems using the LASA handwriting dataset \cite{khansari2011learning}, which comprises a series of 2-D handwritten shapes. Each type comprises 7 demonstrations; 5 of each were selected as train set and the remaining 2 were used as validation set. We sampled all motions with a step size $\Delta t$ of \unit[0.1]{s.} and we trained a distinct discrete-time Koopman system to fit the data for each different shape. 
The encoder and decoder neural networks fully-connected networks with hidden layer sizes of $[50,50,50]$ each and a lifted Koopman dimension of size $20$.
The loss function weights were set to $\lambda=1$, $\mu=0.1$, $\nu=1$, and we set the CBF constant to $\alpha=1$.
We implemented our Koopman learning framework in Pytorch, using the Adam optimizer and a learning rate of $10^{-3}$ on an NVIDIA RTX A5000, and we solved the QP optimization problems using the GUROBI software \cite{gurobi}.

To evaluate the overall performance of our method, we compared our model (which we name KoopmanQP) with two other recent parameterizations of $\mK$ that guarantee stability, named SKEL \cite{fan2022learning} and SOC \cite{mamakoukas2020learning}. It is important to note that while SOC solves another constrained optimization problem (and therefore PGD is required as well), SKEL parameterizes the set of Schur-stable matrices with an unconstrained parameterization which avoids the need for running an optimization-in-the-loop, which can make the training process considerably faster. However, applying PGD through a nested optimization problem allows for the possibility of introducing additional application-specific constraints, e.g., in cases where certain conditions of the dynamical system are known \emph{a priori}.

\begin{figure}[h]
    \centering
    \includegraphics[width=0.95\columnwidth]{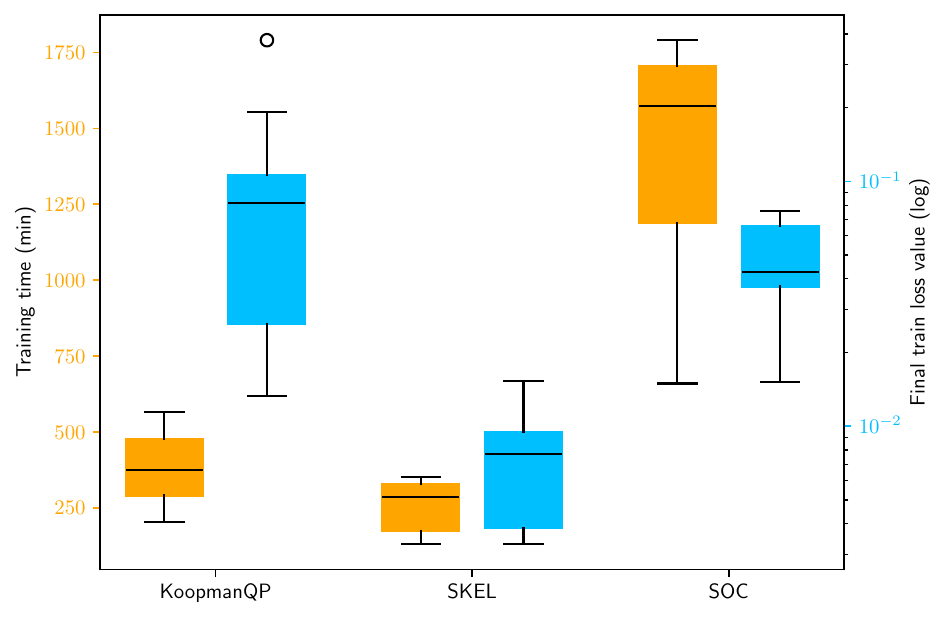}
    \caption{Boxplot comparison of training times (orange) and final training loss value (blue) for the three different architectures.}
    \label{fig:train-loss-time-comparison}
\end{figure}

In a preliminary performance analysis, we compared both the total training time and final train loss value for each model, as presented in \cref{fig:train-loss-time-comparison}. It is evident that including optimization-in-the-loop substantially increases the training time compared to the unconstrained parameterization from SKEL, as expected. Nevertheless, the KoopmanQP architecture still required between 3 to 4 times less training time than the SOC method. It was also unsurprising that the KoopmanQP training loss was the highest among the three methods, due to the additional constraints imposed on the set of admissible learnable matrices. Yet, while the final training losses from SKEL were significantly lower than those of the other methods, the training loss for KoopmanQP was comparable to that of SOC, both being within the same order of magnitude.

\begin{figure*}[htp]
    \centering
    \subfloat[KoopmanQP]{%
        \includegraphics[width=0.31\linewidth]{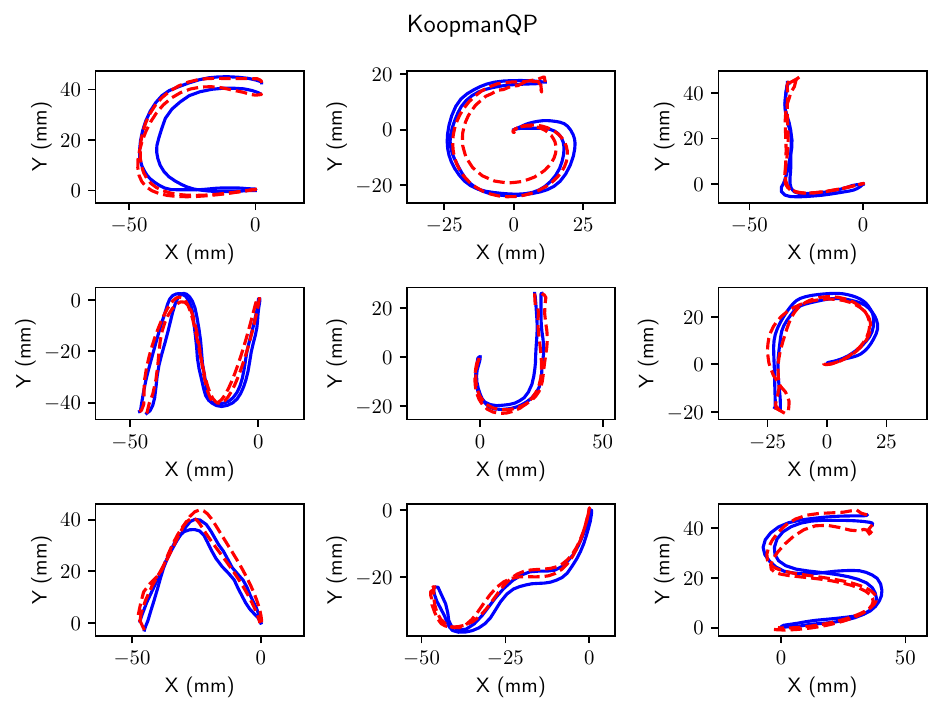}
        \label{fig:koopmanqp-test-results}
    }
    \subfloat[SKEL]{%
        \includegraphics[width=0.31\linewidth]{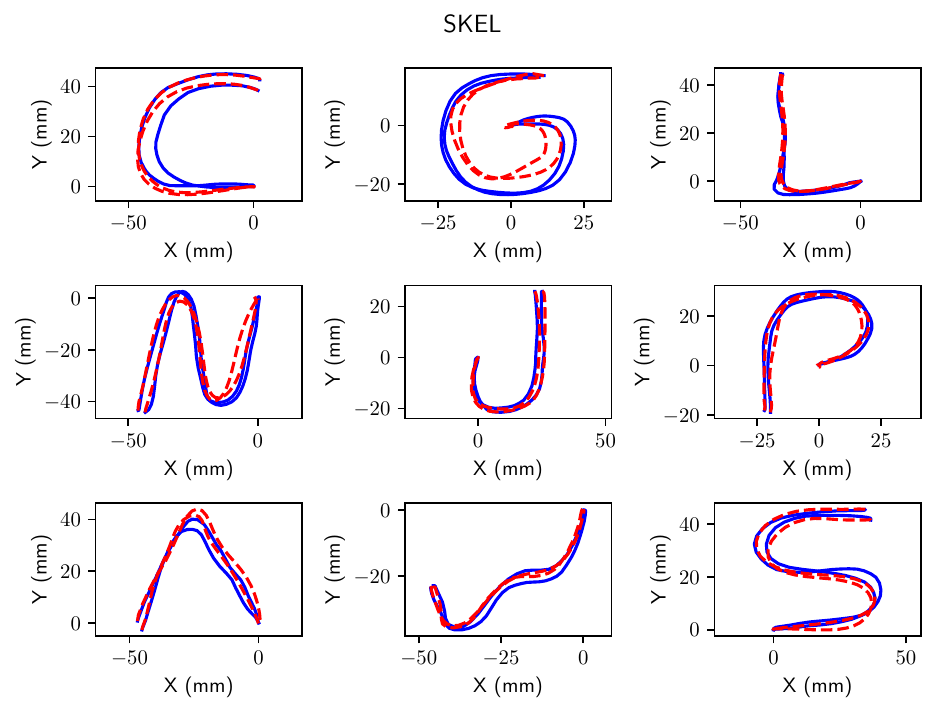}
        \label{fig:skel-test-results}
    }
    \subfloat[SOC]{%
        \includegraphics[width=0.31\linewidth]{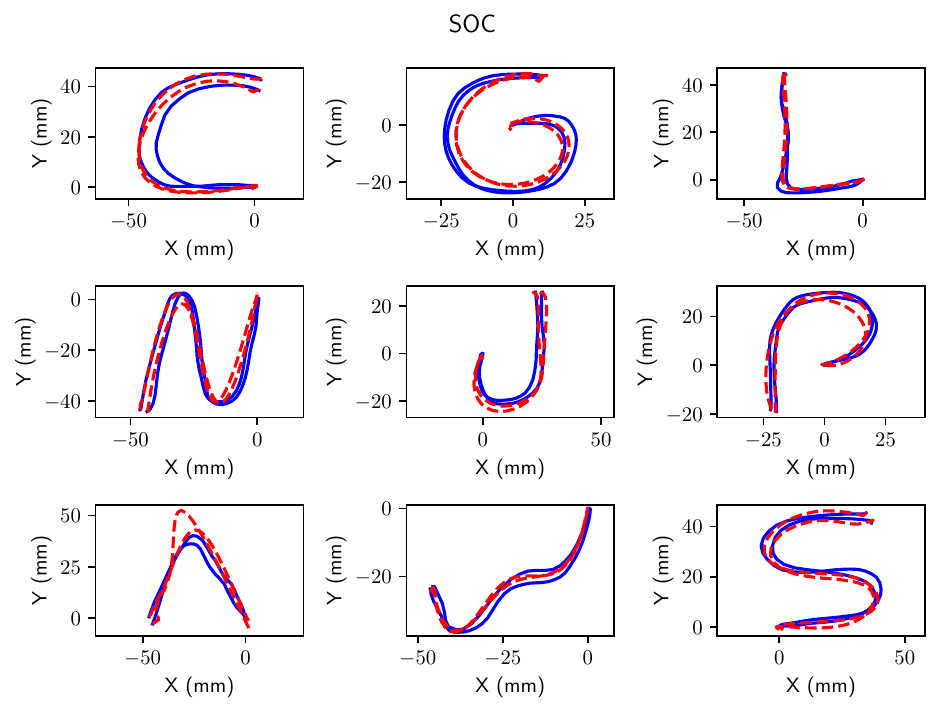}
        \label{fig:soc-test-results}
    }
    \caption{Comparison of the test trajectories for 9 different shapes of the LASA dataset predicted by each one of the 3 compared architectures. In solid blue lines, the ground truth trajectories. In dashed red lines, the predicted Koopman trajectories.}
    \label{fig:test-trajectories}
\end{figure*}

We also evaluated the prediction performance of our model by comparing its normalized mean squared error (NMSE) and standard deviation across the 2 test trajectories of each motion type for each one of the 3 models, which are presented in \cref{tab:metrics-table}. While the NMSE is the lowest in the SKEL method, our model achieves the lowest normalized standard deviation values, despite showing the largest NMSE value. A probable explanation for this result is the fact that our stability constraints are sufficient but not necessary, as opposed to the parameterization of the other two architectures.
\Cref{fig:test-trajectories} shows a model comparison of the test trajectories for 9 different shapes.

\begin{table}[h]
\centering
\renewcommand{\arraystretch}{1.5} 
\begin{tabular}{|c|ccc|}
\hline
\rowcolor[HTML]{C0C0C0} 
\cellcolor[HTML]{C0C0C0} & \multicolumn{3}{c|}{\cellcolor[HTML]{C0C0C0}\textbf{Model type}} \\ \cline{2-4} 
\rowcolor[HTML]{C0C0C0} 
\multirow{-2}{*}{\cellcolor[HTML]{C0C0C0}\textbf{}} &
  \multicolumn{1}{c|}{\cellcolor[HTML]{C0C0C0}\textbf{KoopmanQP}} &
  \multicolumn{1}{c|}{\cellcolor[HTML]{C0C0C0}\textbf{SKEL}} &
  \multicolumn{1}{c|}{\cellcolor[HTML]{C0C0C0}\textbf{SOC}} \\ \hline
NMSE (mm)                     & \multicolumn{1}{c|}{$0.17$}       & \multicolumn{1}{c|}{$\mathbf{0.11}$}      & \multicolumn{1}{c|}{$0.12$} \\ \hline
NormSTD (mm)           & \multicolumn{1}{c|}{$\mathbf{9.18\cdot 10^{-2}}$} & \multicolumn{1}{c|}{$1.01\cdot 10^{-1}$} & \multicolumn{1}{c|}{$9.26\cdot10^{-2}$} \\ \hline
\end{tabular}
\caption{Normalized mean squared error and standard deviation across all test trajectories for the 3 analyzed models.}
\label{tab:metrics-table}
\end{table}

\section{CONCLUSIONS AND FUTURE WORK}

In this paper we presented a new sufficient condition for the stability of discrete-time systems that leads to a series of piecewise linear constraints which can be decoupled by rows of the system matrix, making them suitable to be included into quadratic programming optimization problems. We explored the utility of this method by employing it to impose asymptotic stability during the joint learning of a Koopman system and its corresponding observables, parameterized by neural networks. In particular, we applied a control barrier function-based projected gradient descent after each update step to enforce gradual updates of the parameter towards the feasible set, but while preventing abrupt undesired updates. 

We evaluated the performance of our system on the LASA Handwriting dataset, consisting of a series of distinct 2-D hand motions, and compared the results of the trained models with two other recent parameterizations of the Koopman matrix that guarantee stability. We showed how, even though our proposed sufficient condition does not span all the Schur stable matrices, we obtained a similar prediction performance compared to the two other methods. Moreover, the training time of our optimization method greatly outperformed the training time shown by the SOC method, which also relies on solving an optimization problem within the training loop.

Regarding future work, we plan to extend our research to controlled systems, and study how learning nonlinear stable closed-loop dynamics using the Koopman operator can facilitate the efficient computation of nonlinear control laws. Moreover, since the results did not show a significant decrease in performance despite the more restrictive stability condition, we plan to study how we can further reduce the computation time without substantially compromising the prediction performance. Finally, we want to take advantage of the flexibility introduced by the optimization problem, and explore the integration of additional observables that represent physical quantities of the system with certain constraints (e.g., an energy conservation law) that can be enforced by the optimization problem.

\bibliographystyle{IEEEtran} 

\bibliography{biblio/IEEEfull,biblio/IEEEConfFull,biblio/OtherFull,
  biblio/tron,%
  biblio/formationControl,%
  biblio/websites,%
  biblio/koopmanQP,%
  biblio/koopman,%
  biblio/math,%
  biblio/koopmanControl}

\end{document}